\documentclass{ws-procs9x6_mg}
\usepackage{epsfig}

\begin{document}

\title{
Jets at CDF\footnote{\uppercase{P}resented at the 
workshop on 
``\uppercase{L}ow-x \uppercase{P}hysics'', 
\uppercase{L}isbon, \uppercase{P}ortugal, \uppercase{J}une 28-\uppercase{J}uly 1, 2006}}
\author{Michele Gallinaro\footnote{\uppercase{R}epresenting the \uppercase{CDF} collaboration.}}

\address{Laboratory of Experimental High Energy Physics\\
The Rockefeller University,\\
1230 York Avenue, New York, NY 10021, USA}

\maketitle

\abstracts{
Recent jet results in $p\bar{p}$ collisions at $\sqrt{s}$=1.96~TeV from the CDF experiment at the Tevatron are presented.
The jet inclusive cross section is compared to next-to-leading order QCD prediction in different rapidity regions.
The $b$-jet inclusive cross section is measured exploiting the long lifetime and large mass of $B$-hadrons.
Jet shapes, W+jets and W/Z+photon cross sections are also measured and compared to expectations from QCD production.
}

\section{Introduction}

The current physics program at the Tevatron hadron collider includes studies of jets, with the goal of performing precision measurements to test 
and further constrain the validity of the Standard Model.
Jets are collimated sprays of particles originating from the fragmentation of the initial interacting partons.
The full jet reconstruction allows to measure the primary parton's energy.
Jet energies are measured experimentally by adding the energy of the calorimeter cells associated to a cluster, using predetermined algorithms.
During Run~II, both the CDF and the D\O~collaborations are studying alternative methods to the cone-based JetClu algorithm\cite{JetClu} used during Run~I (1992-1995), 
in order to avoid problems of infrared and collinear divergences due to soft partons and below/above threshold particle emission.
Both MidPoint\cite{midpoint} and $k_T$\cite{kt} jet reconstruction algorithms have been used in Run~II.
The former is an improved version of the 
JetClu algorithm which reduces the sensitivity to infrared and collinear problems.
The latter starts by finding pairs of nearby particles 
and merges them together to form new pseudo-particles,
continuing until a set of stable well-separated jets are found, and is infrared and collinear safe to all orders in perturbative QCD (pQCD).
In addition to the energy from the primary parton, jets accrue soft contributions from the underlying event (UE) of beam remnants. 
These contributions become more important at smaller jet $p_T$.
During Run~I and Run~II, the contribution from UE energy has been studied and a modification 
to Pythia\cite{pythia} MC has been determined (Tune A\cite{tuneA}) using CDF data. 
Pythia MC with the new set of parameters describes well the jet shapes measured in Run~II\cite{jetshapes}.

\section{Jet cross section}
\begin{figure}[tp]
\epsfxsize=1.0\textwidth
\centerline{\epsfig{figure=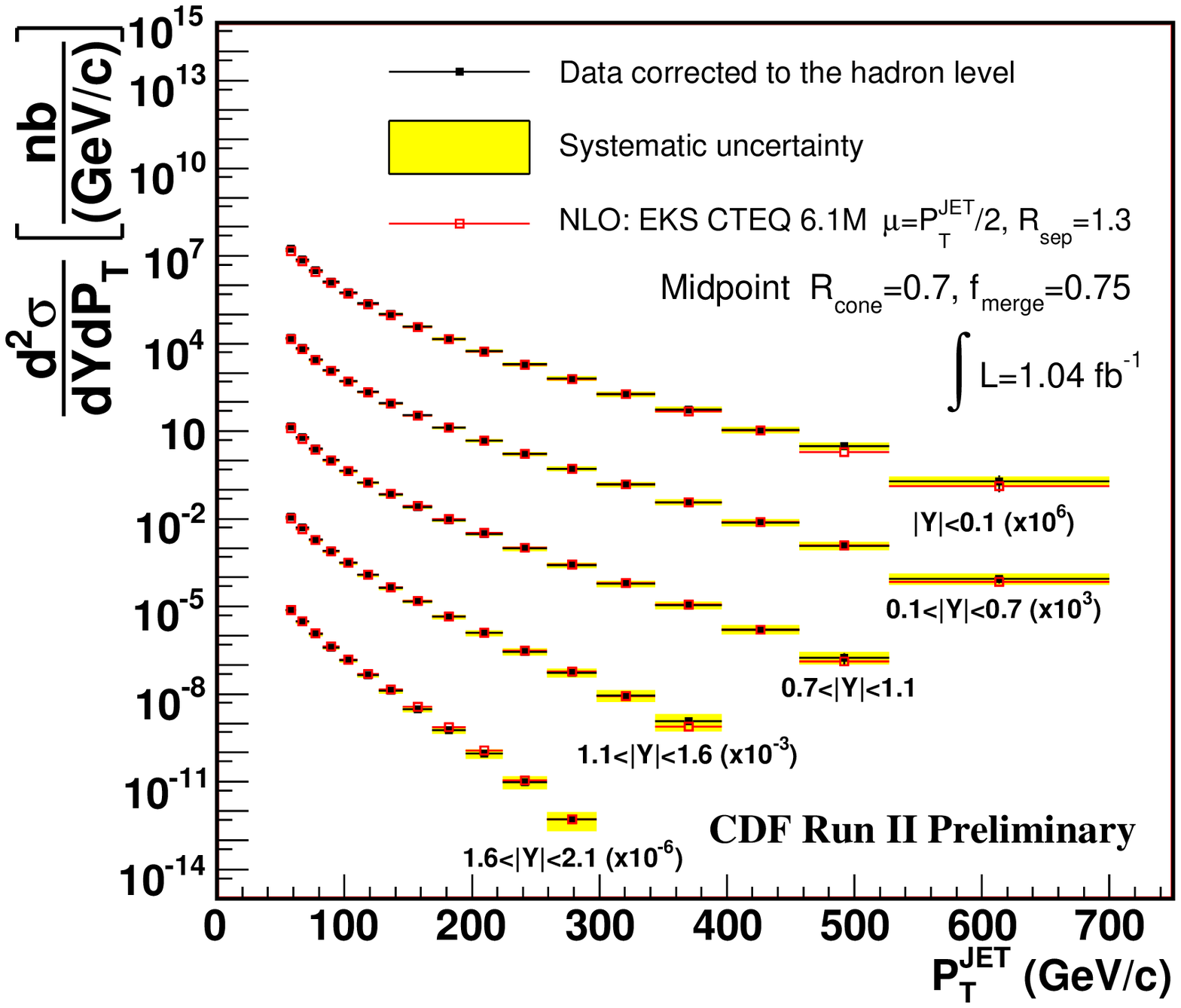,width=0.53\hsize} \hspace*{-0.3cm}
            \epsfig{figure=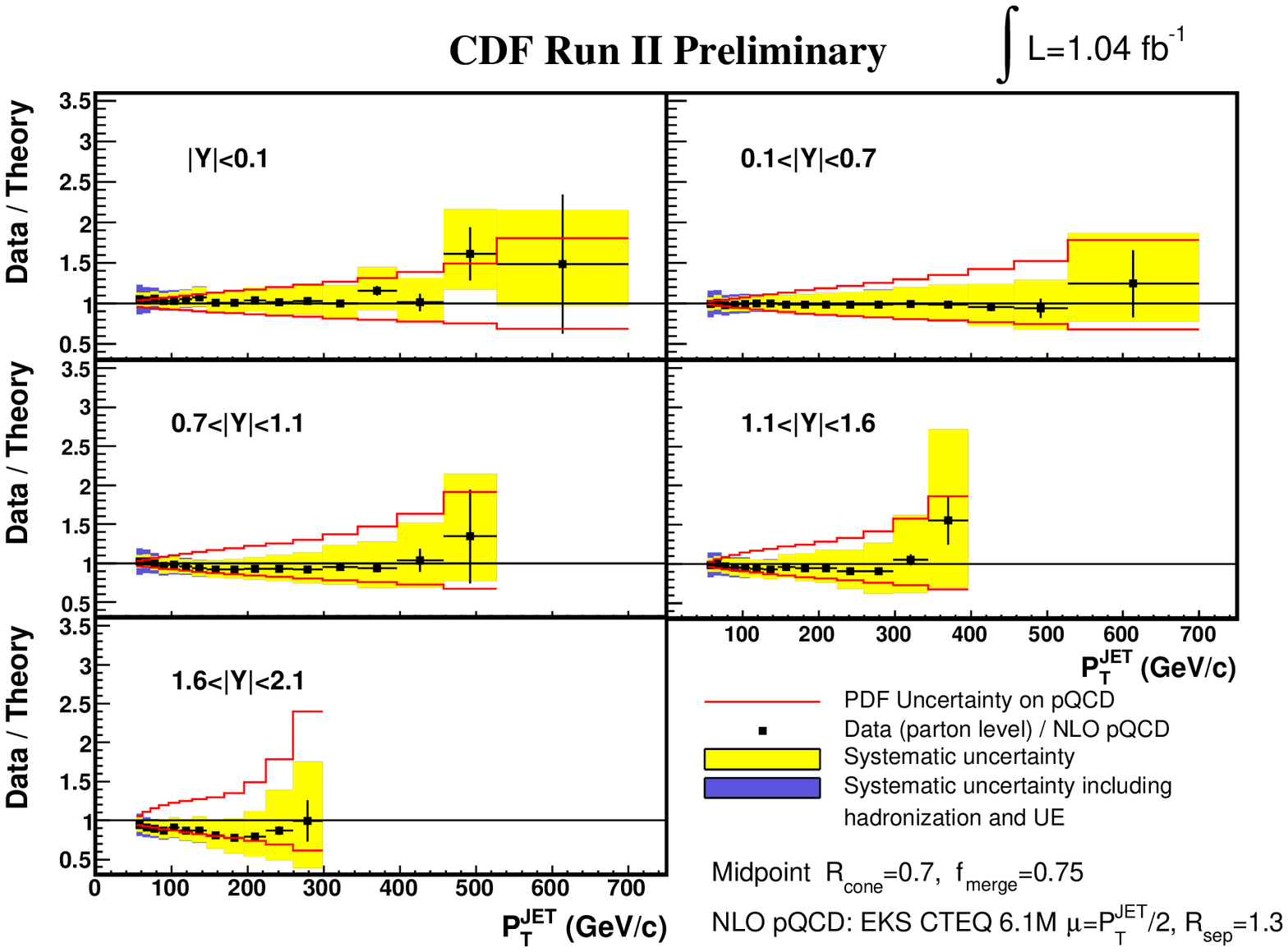,width=0.57\hsize}}
\caption{\label{fig:jetxsec}
{\em Left}: Measured inclusive jet cross section (circles) as a function of jet $p_T$ compared to NLO pQCD predictions (histogram).
The shaded band shows the total systematic uncertainty of the measurement.
{\em Right}: Ratio Data/Theory as a function of jet $p_T$. The error bar (shaded area) shows the statistical (systematical) uncertainty.
The solid lines indicate the PDF theoretical uncertainty.}
\end{figure}

Measurement of the jet inclusive cross section provides a powerful test of pQCD and
a sensitive probe of quark sub-structure down to a scale of $\approx10^{-17}$cm.
Run~I data exhibited an excess in jet cross section for events with high $E_T$ jets\cite{xsec_run1}, when 
compared to the then current {\it Parton Distribution Functions} (PDFs).
The increase in center-of-mass energy from 1.80 to 1.96~TeV
from Run~I to Run~II results in a larger kinematic range for measuring jet production.
With a data sample ten times larger than in Run~I, jet production can be 
measured at transverse energies far beyond those of Run~I and spanning more than eight orders of magnitude.
The measurement of the inclusive jet production cross section is performed using jets clustered by the MidPoint cone algorithm 
for jets with $p_T\ge 54$~GeV/c in five different jet rapidity regions: 
$|y|<0.1$, $0.1<|y|<0.7$, $0.7<|y|<1.1$, $1.1<|y|<1.6$, and $1.6<|y|<2.1$ (Fig.~\ref{fig:jetxsec}, left). 
Results are based on over 1 fb$^{-1}$ of Run~II data. 
The measured cross section is in agreement with next-to-leading order (NLO) pQCD after the necessary non-perturbative parton-to-hadron corrections are taken into account.
Figure~\ref{fig:jetxsec} (right) shows the ratio of Data/Theory as a function of jet $p_T$, and good agreement is observed in the whole range. 
It also shows the ratio using MRST2004 and CTEQ6.1M PDFs,
indicating that the current measurement in the forward region has the power of reducing the PDF uncertainties.
Comparison of results in the central and forward rapidity regions can be used as a powerful discriminant 
to select new physics processes, free of common systematic uncertainties.
The jet cross section has also been measured using the $k_T$ algorithm yielding similar results\cite{jetxsec_kt}.

\section{Heavy flavor}

\begin{figure}[tp]
\epsfxsize=1.0\textwidth
\centerline{\epsfig{figure=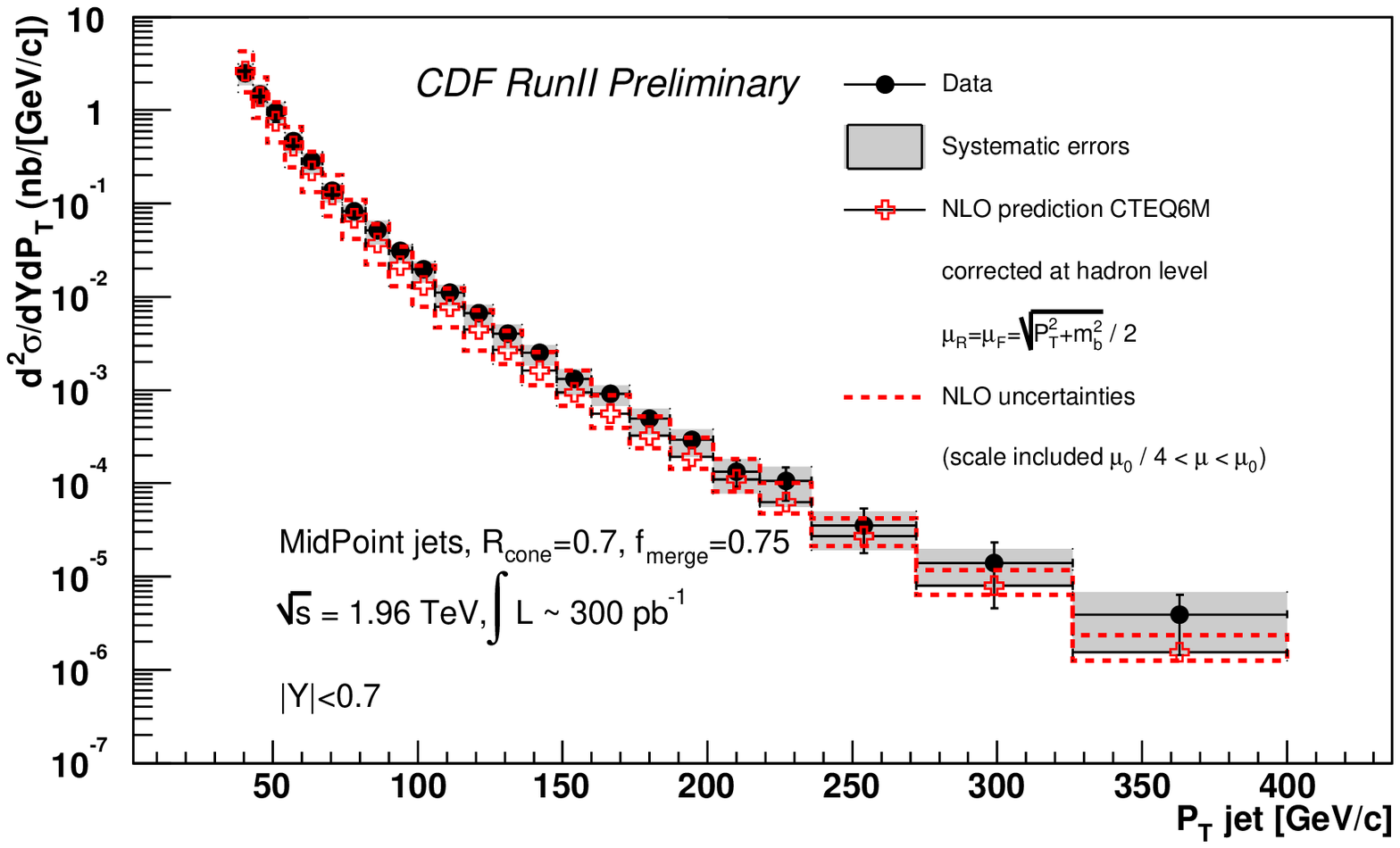,width=0.57\hsize} \hspace*{-0.3cm}
            \epsfig{figure=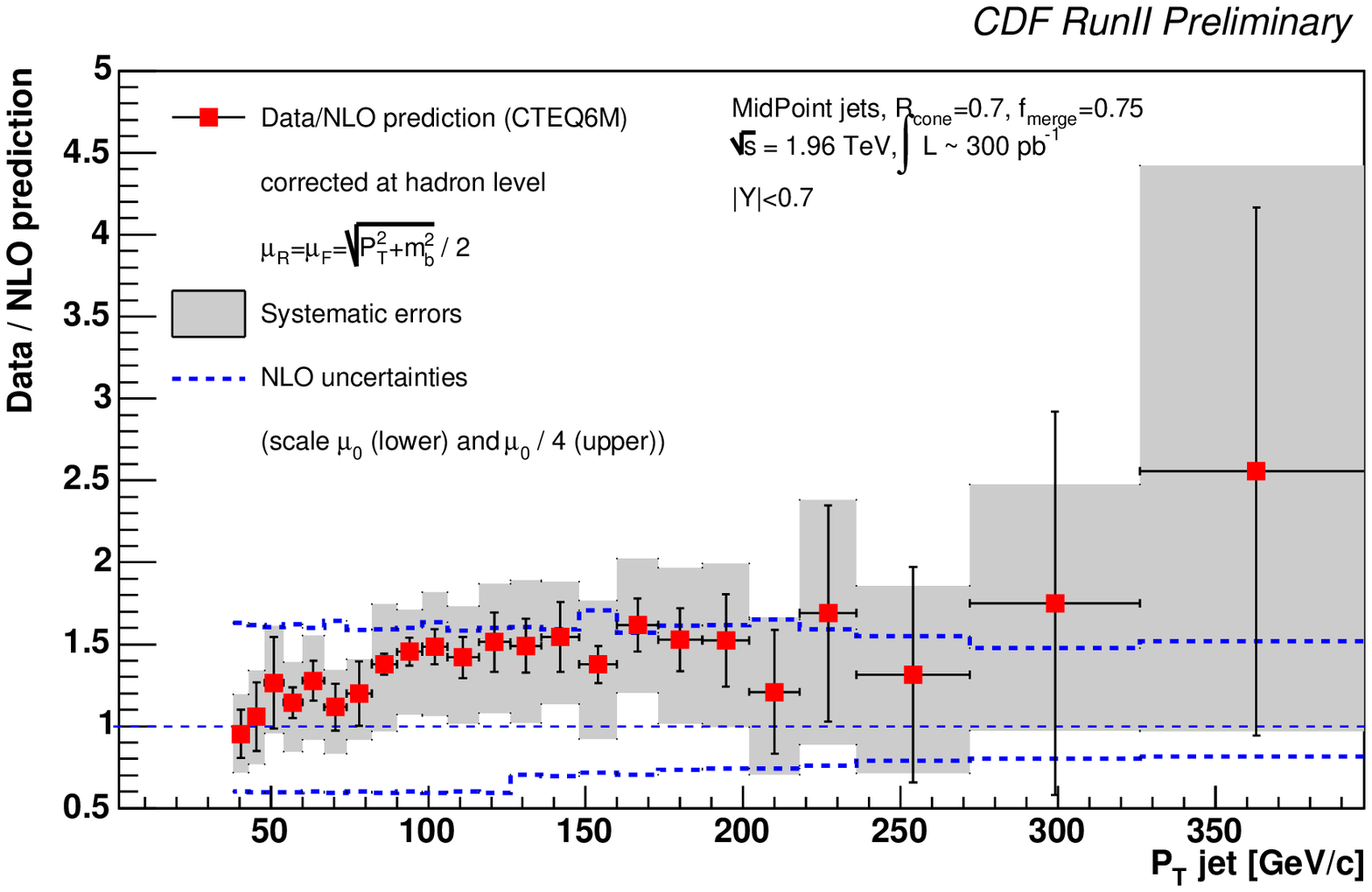,width=0.57\hsize}}
\caption{\label{fig:bjetxsec}
{\em Left}: Measured inclusive $b-$jet cross section (circles) as a function of jet $p_T$ compared to NLO pQCD predictions (histogram).
The shaded band shows the total systematic uncertainty of the measurement.
{\em Right}: Ratio of measured to NLO cross section as a function of jet $p_T$. 
The error bar (shaded area) shows the statistical (systematical) uncertainty.}
\end{figure}

The inclusive $b-$jet production cross section has been measured using 300~pb$^{-1}$. 
The measurement is performed for jets reconstructed using the MidPoint algorithm with transverse 
momentum in the range $38<p_T^{jet} < 400$~GeV/c and rapidity $| y | < 0.7$.
It exploits the good tracking capabilities of the CDF detector to tag $b-$jets by reconstructing secondary vertices.
The $b-$tagging algorithm uses displaced tracks associated with a jet and within a cone 
of $\Delta R=$0.4 in the $\eta-\phi$ space with respect to the jet axis.
Secondary vertices are searched for using the significance of the impact parameter 
and the displacement of the secondary vertex from the primary interaction, L2D. 
The sign of L2D is used to distinguish heavy flavor candidates from mis-tagged jets, 
and only positively tagged jets (L2D$>$0) are considered. 
The $b-$tagging algorithm is optimized for a jet cone of $R=0.4$ in order to avoid an increase in the mis-tagging rate.
Both $B$ and $D$ hadrons have a displaced measured secondary vertex.
Using Pythia MC, the shape of the invariant mass distribution of all charged tracks attached to the secondary vertex is used to extract the $b-$jet fraction.
The measured cross section is in agreement with NLO pQCD predictions (Fig.~\ref{fig:bjetxsec}, left).
Figure ~\ref{fig:bjetxsec} (right) shows the ratio ($R$) between measured and theoretical predictions, and the grey band represents the total systematic error from data. 
Systematic theoretical uncertainties are shown around $R=1$. For $b-$jets with transverse momentum below 90~GeV/c, 
good consistency between the measured cross section and NLO pQCD prediction is found; 
for $b-$jets with $p_T> 90$~GeV/c, agreement is observed within the systematic uncertainties.

\section{W+jet production}

\begin{figure}[tp]
\epsfxsize=1.0\textwidth
\centerline{\epsfig{figure=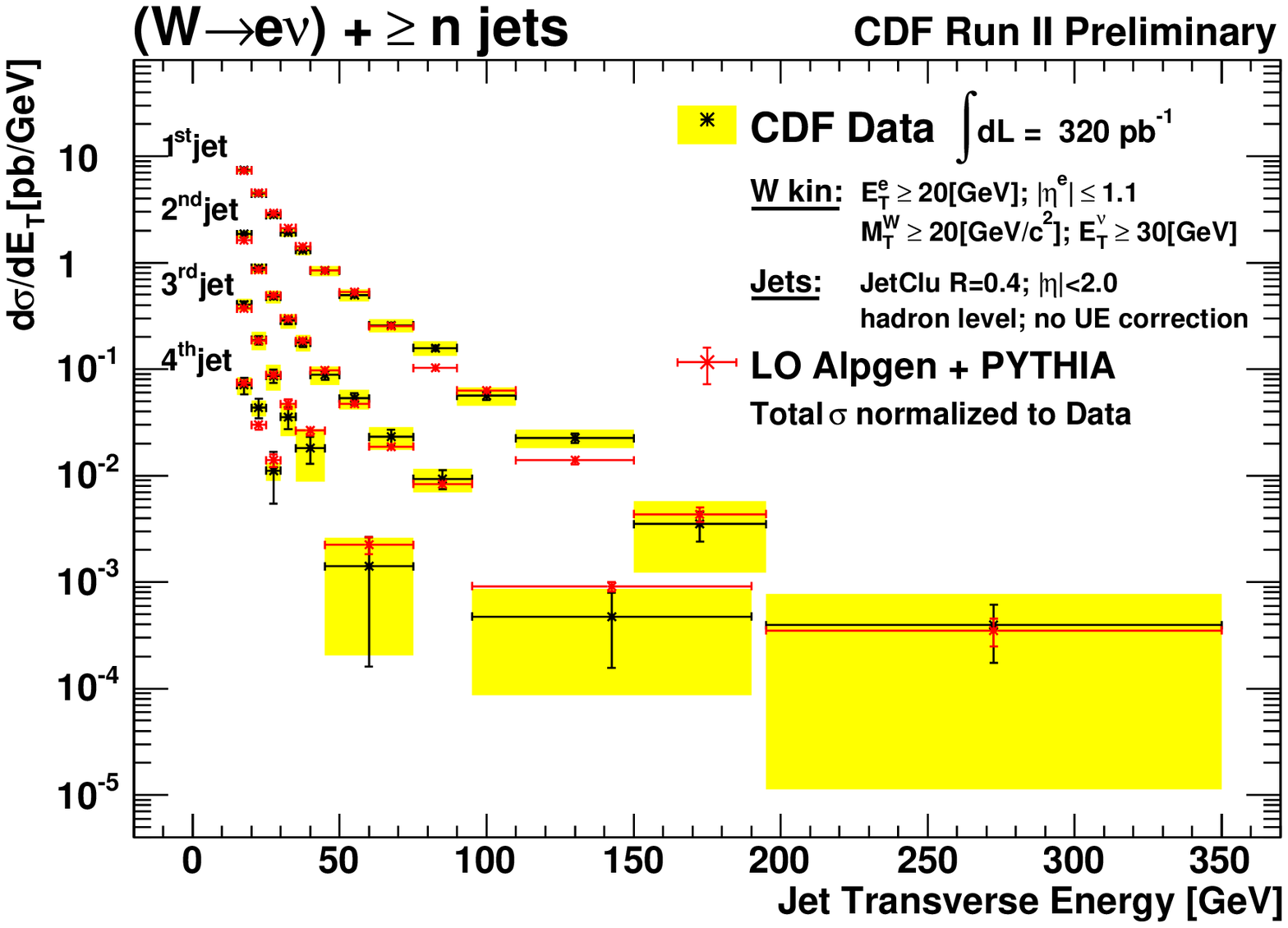,width=0.57\hsize} \hspace*{-0.3cm}
            \epsfig{figure=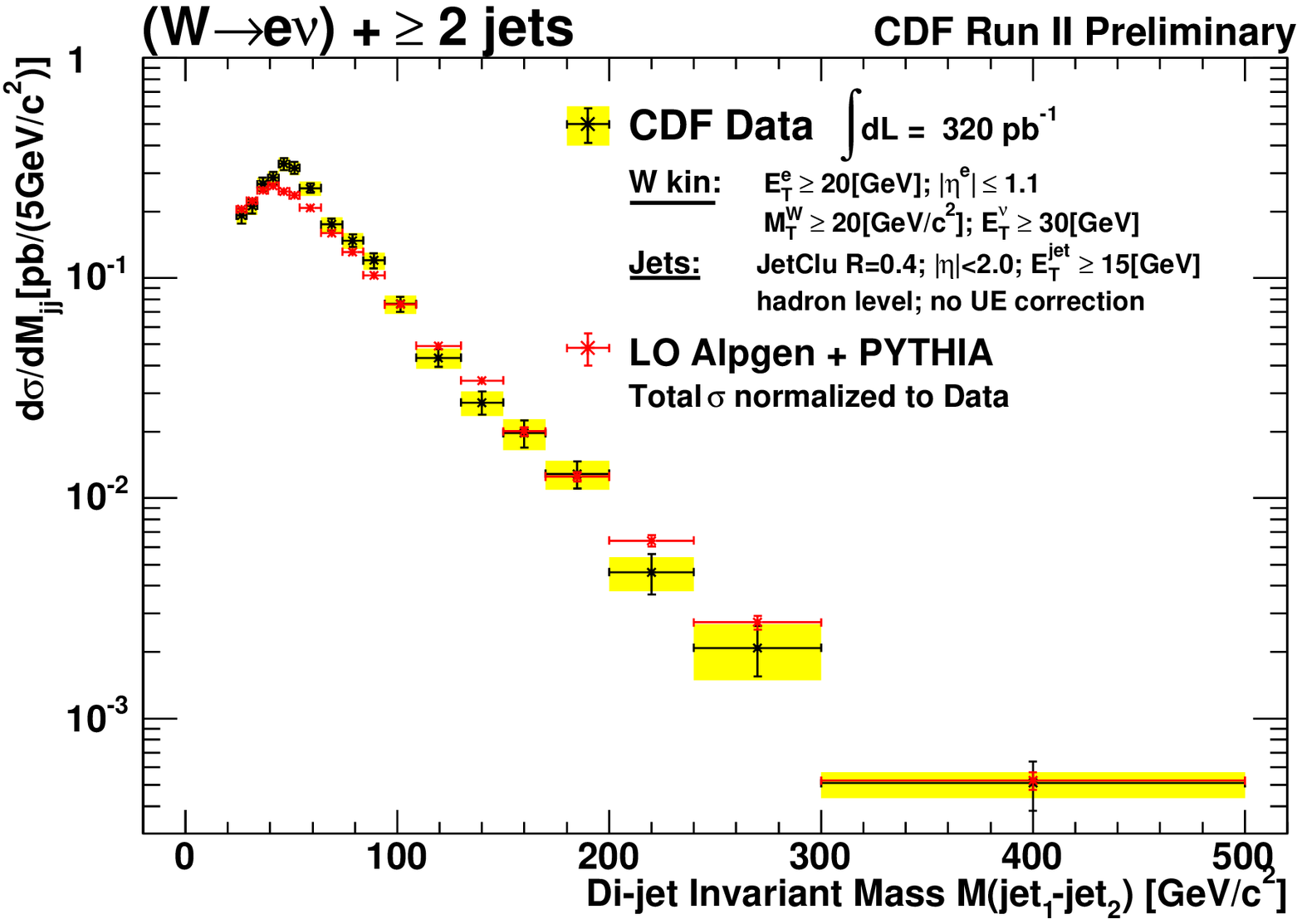,width=0.57\hsize}}
\caption{\label{fig:wjet}
{\em Left}: Differential cross section $d\sigma (W\rightarrow e\nu+ \ge n~{\rm jets}) /dE_T^{jet}$ as a function of the jet transverse energy. 
Data are compared to Alpgen+Pythia MC and normalized to the measured inclusive cross section in each jet multiplicity sample.
{\em Right}: differential cross section as a function of the dijet invariant mass.}
\end{figure}

The study of inclusive $W$ boson production in association with jets provides another useful test of QCD in a large $Q^2$ environment.
The measurement of $W+$jet production cross section may also improve the understanding of the background 
to top pair production and possibly increase the sensitivity to Higgs boson searches.
Recent work has been focused on improving the Monte Carlo simulation with multi-particle production in the final state.
A new measurement of $W+n$~jet ($n\ge 1,...4$) cross section as a function of jet $p_T$ has been performed (Fig.~\ref{fig:wjet}, left). 
Cross sections have been corrected to particle level jets, in the effort of minimizing the detector effects and reducing the dependence from a particular model. 
Data show good agreement with the Alpgen\cite{alpgen}+Pythia MC over the entire jet $p_T$ spectrum.
Other kinematic variables studied indicate good agreement between data and MC predictions. 
Figure~\ref{fig:wjet} (right) shows the measured differential cross section as a function of the dijet invariant mass.

\section{W/Z+photon production}

The study of dijet mass resolution is very important as many new physics signatures are expected to appear as dijet mass bumps\cite{Bocci_etal}. 
One example is the neutral Higgs boson which is expected to decay predominantly to $b \bar b$, 
and the ability to precisely reconstruct the dijet mass peak can thus enhance the Higgs discovery potential.
A study of the process $p\bar{p}\rightarrow W(Z)\gamma\rightarrow q\bar{q}\gamma$
has been performed using $\approx 184$~pb$^{-1}$ of CDF data\cite{tesiBocci}.
A dedicated trigger was studied and implemented to enhance the sensitivity of this analysis.
The event sample is selected by requiring a central photon with $E_T>12$~GeV/c and two jets with $E_T>15$~GeV/c.
Thanks to the low threshold of the photon transverse energy, the $W/Z$ mass peak is not biased and the background is estimated from data.
An advanced Neural Network event selection improves the signal-to-noise ratio from 1/333 to 1/71.
Further optimization around the $W/Z$ dijet mass region increases the ratio to 1/41. 
No evidence of a $W/Z\rightarrow q\bar{q}$ peak is found, 
and the 95\% upper limit of the production cross section $\sigma(p\bar{p}\rightarrow W\gamma)$ is calculated to be 54~pb, 
consistent with the Standard Model expectations of 20.5~pb.

\end{document}